# Functional Pearl: Theorem Proving for All

## Equational Reasoning in Liquid Haskell

Niki Vazou
University of Maryland, USA

Joachim Breitner
University of Pennsylvania, USA

Will Kunkel
University of Maryland, USA

David Van Horn
University of Maryland, USA

Graham Hutton
University of Nottingham, UK

## Abstract
Equational reasoning is one of the key features of pure functional languages such as Haskell. To date, however, such reasoning always took place externally to Haskell, either manually on paper, or mechanised in a theorem prover. This article shows how equational reasoning can be performed directly and seamlessly within Haskell itself, and be checked using Liquid Haskell. In particular, language learners — to whom external theorem provers are out of reach — can benefit from having their proofs mechanically checked. Concretely, we show how the equational proofs and derivations from Hutton [2016]'s textbook can be recast as proofs in Haskell (spoiler: they look essentially the same).

## 1 Introduction

Advocates for pure functional languages such as Haskell have long argued that a key benefit of these languages is the ability to reason equationally, using basic mathematics to reason about, verify, and derive programs. Consequently, introductory textbooks often place considerable emphasis on the use of equational reasoning to prove properties of programs. Hutton [2016], for example, concludes the chapter on equational reasoning in *Programming in Haskell* with the remark, "Mathematics is an excellent tool for guiding the development of efficient programs with simple proofs!"

In this pearl, we mechanize equational reasoning in Haskell using an expressive type system. In particular, we demonstrate how Liquid Haskell [Vazou 2016], which brings refinement types to Haskell, can effectively check pen-and-paper proofs. Doing so remains faithful to the traditional techniques for verifying and deriving programs, while enjoying the added benefit of being mechanically checked for correctness. Moreover, this approach is well-suited to beginners because the language of proofs is simply Haskell itself.

To demonstrate the applicability of our approach, we present a series of examples[1] that replay equational reasoning, program optimizations, and program calculations from the literature. The paper is structured as follows:

- **Equational Reasoning (§ 2):** We prove properties about familiar functions on lists, and compare them with standard textbook proofs. In each case, the proofs are strikingly similar! This approach opens up machine-checked equational reasoning to ordinary Haskell users, without requiring expertise in theorem provers.

- **Optimized Function Derivations (§ 4):** Another common theme in Haskell textbooks is to derive efficient implementations from high-level specifications, as described by Bird [1987, 2010] and Hutton [2016]. We demonstrate how Liquid Haskell supports the derivation of correct-by-construction programs by using equational proofs that themselves serve as efficient implementations.

- **Calculating Correct Compilers (§ 5):** As an extended case study we use equational reasoning to derive a correct-by-construction compiler for a simple language of numeric expressions, following Hutton [2016].

- **Related Work (§ 6):** Finally, we compare how this style of equational reasoning relates to proofs in other theorem provers and programming languages.

We conclude that, even though these proofs can be performed in external tools such as Agda, Coq, Dafny, Isabelle or Lean, equational reasoning using Liquid Haskell is unique in that the proofs are literally just Haskell functions. It can therefore be used by any Haskell programmer or learner.

## 2 Reasoning about Programs

The goal of Liquid Haskell [Vazou 2016] is to move reasoning about Haskell programs into the programs themselves and to automate this reasoning as much as possible. It accomplishes this goal by extending the Haskell language with refinement types [Freeman and Pfenning 1991], which are checked by an external SMT solver [Barrett et al. 2010].

### 2.1 Lightweight Reasoning

The power of SMT solving allows Liquid Haskell to prove certain properties entirely automatically, with no user input; we call these *lightweight program properties*.

***Linear Arithmetic***   Many properties involving arithmetic can be proved automatically in this manner. For example, given the standard length function on lists

```
length :: [a] → Int
```

---

[1] All examples can be found in the interactive, browser-based demo at http://goto.ucsd.edu/~nvazou/theorem-proving-for-all/; source code at https://github.com/nikivazou/EquationalReasoningInLiquidHaskell.

```
length []     = 0
length (_:xs) = 1 + length xs
```

we might find it useful to specify that the length of a list is never negative. Liquid Haskell extends the syntax of Haskell by interpreting comments of the form `{-@ ... @-}` as declarations, which we can use to express this property:

```
{-@ length :: [a] → {v:Int | 0 <= v } @-}
```

Liquid Haskell is able to verify this specification automatically due to the standard refinement typing checking [Vazou et al. 2014] automated by the SMT solver:

- In the first equation in the definition for `length`, the value $v$ is 0, so the SMT solver determines that $0 \leq v$.
- In the second equation, the value $v$ is $1 + v'$, where $v'$ is the result of the recursive call to `length xs`. From the refinement type of `length`, Liquid Haskell knows $0 \leq v'$, and the SMT solver can deduce that $0 \leq v$.

Proving that the length of a list is non-negative is thus fully automated by the SMT solver. This is because SMT solvers can efficiently decide linear arithmetic queries, so verifying this kind of property is tractable. Note that the refinement type does not mention the recursive function `length`.

*Measures* In order to allow Haskell functions to appear in refinement types, we need to lift them to the refinement type level. Liquid Haskell provides a simple mechanism for performing this lifting on a particular restricted set of functions, called *measures*. Measures are functions which: take one parameter, which must be an algebraic data type; are defined by a single equation for each constructor; and in their body call only primitive (*e.g.,* arithmetic) functions and measures. For this restricted class of functions, refinement types can still be checked fully automatically.

For instance, `length` is a measure: it has one argument, is defined by one equation for each constructor, and calls only itself and the arithmetic operator `(+)`. To allow `length` to appear in refinements, we declare it to be a measure:

```
{-@ measure length @-}
```

For example, we can now state that the length of two lists appended together is the sum of their lengths:

```
{-@ (++) :: xs:[a] → ys:[a] → {zs:[a] |
      length zs == length xs + length ys} @-}

(++) :: [a] → [a] → [a]
[]     ++ ys = ys
(x:xs) ++ ys = x : (xs ++ ys)
```

Liquid Haskell checks this refinement type in two steps:

- In the first equation in the definition of `(++)`, the list `xs` is empty, thus its length is 0, and the SMT solver can discharge this case via linear arithmetic.

- In the second equation case, the input list is known to be `x:xs`, thus its length is `1 + length xs`. The recursive call additionally indicates that `length (xs ++ ys) = length xs + length ys` and the SMT solver can also discharge this case using linear arithmetic.

### 2.2 Deep Reasoning

We saw that because `length` is a measure, it can be lifted to the refinement type level while retaining completely automatic reasoning. We cannot expect this for recursive functions in general, as quantifier instantiation leads to unpredictable performance [Leino and Pit-Claudel 2016].

The append function, for example, takes two arguments, and therefore is not a measure. If we lift it to the refinement type level, the SMT solver will not be able to automatically check refinements involving it. Liquid Haskell still allows reasoning about such functions, but this limitation means the user may have to supply the proofs themselves. We call properties that the SMT solver cannot solve entirely automatically *deep program properties*.

For example, consider the following definition for the reverse function on lists in terms of the append function:

```
{-@ reverse :: is:[a] →
      {os:[a] | length is == length os} @-}

reverse :: [a] → [a]
reverse []     = []
reverse (x:xs) = reverse xs ++ [x]
```

Because the definition uses append, which is not a measure, the reverse function itself is not a measure, so reasoning about it will not be fully automatic.

In such cases, Liquid Haskell can lift arbitrary Haskell functions into the refinement type level via the notion of *reflection* [Vazou et al. 2018]. Rather than using the straightforward translation available for measures, which completely describes the function to the SMT solver, reflection gives the SMT solver only the value of the function for the arguments on which it is actually called. Restricting the information available to the SMT solver in this way ensures that checking refinement types remains decidable.

To see this in action, we prove that reversing a singleton list does not change it, *i.e.,* `reverse [x] == [x]`. We first declare reverse and the append function as reflected:

```
{-@ reflect reverse @-}
{-@ reflect ++      @-}
```

We then introduce the function `singletonP`, whose refinement type expresses the desired result, and whose body provides the proof in equational reasoning style:

```
{-@ singletonP :: x:a →
      {reverse [x] == [x]} @-}

singletonP :: a → Proof
```

```
singletonP x
  =   reverse [x]
      -- applying reverse on [x]
  ==. reverse [] ++ [x]
      -- applying reverse on []
  ==. [] ++ [x]
      -- applying ++ on [] and [x]
  ==. [x]
  *** QED
```

We can understand this function as mapping a value x to a proof of reverse [x] == [x]. The type Proof is simply Haskell's unit type (), and {reverse [x] == x} is syntactic sugar for {v:() | reverse [x] == x}, a refinement of the unit type. This syntax hides the irrelevant value v :: ().

Note that the body of the singletonP function looks very much like a typical pen-and-paper proof, such as the one in Hutton [2016]'s book. The correspondence is so close that we claim proving a property in Liquid Haskell can be just as easy as proving it on paper by equational reasoning — but the proof in Liquid Haskell is machine-checked!

As always, Liquid Haskell uses an SMT solver to check this proof. Because the body of singletonP syntactically contains the three terms reverse [x], reverse [] and [] ++ [x], Liquid Haskell passes the corresponding equations

```
reverse [x] = reverse [] ++ [x]
reverse []  = []
[] ++ [x]   = [x]
```

to the SMT solver, which then easily derives the desired property reverse [x] = [x]. Note that the following definition would constitute an equally valid proof:

```
singletonP x =
  const () (reverse [x], reverse [], [] ++ [x])
```

But such a compressed "proof" is neither easy to come up with directly, nor is it readable or very insightful. Therefore, we use proof combinators to write readable equational-style proofs, where each reasoning step is checked.

***Proof Combinators***  As already noted in the previous section, we use Haskell's unit type to represent a proof:

```
type Proof = ()
```

The unit type is sufficient because a theorem is expressed as a refinement on the arguments of a function. In other words, the "value" of a theorem has no meaning.

Proof combinators themselves are simply Haskell functions, defined in the Equational[2] module that comes with Liquid Haskell. The most basic example is ***, which takes any expression and ignores its value, returning a proof:

```
data QED = QED

(***) :: a → QED → Proof
```

---
[2]See the Equational.hs module on github.

```
_ *** QED = ()
infixl 2 ***
```

The QED argument serves a purely aesthetic purpose, allowing us to conclude proofs with *** QED.

***Equational Reasoning***  The key combinator for equational reasoning is the operator (==.). Its refinement type ensures its arguments are equal, and it returns its second argument, so that multiple uses of (==.) can be chained together:

```
{-@ (==.) :: x:a → y:{a | x == y} →
        {o:a | o == y && o == x} @-}

(==.) :: a → a → a
_ ==. x = x
```

***Explanations***  Sometimes we need to refer to other theorems in our proofs. Because theorems are just Haskell functions, all we need is an operator that accepts an argument of type Proof, which is defined as follows:

```
(?) :: a → Proof → a
x ? _ = x
```

For example, we can invoke the theorem singletonP for the value 1 simply by mentioning singletonP 1 in a proof:

```
{-@ singleton1P :: { reverse [1] == [1] } @-}
singleton1P
  =   reverse [1]
  ==. [1] ? singletonP 1
  *** QED
```

Note that although the ? operator is suggestively placed next to the equation that we want to justify, its placement in the proof is actually immaterial — the body of a function equation is checked all at once.

### 2.3 Induction on Lists

Structural induction is a fundamental technique for proving properties of functional programs. For the list type in Haskell, the principle of induction states that to prove that a property holds for all (finite) lists, it suffices to show that:

- It holds for the empty list [] (the base case), and
- It holds for any non-empty list x:xs assuming it holds for the tail xs of the list (the inductive case).

Induction does not require a new proof combinator. Instead, proofs by induction can be expressed as recursive functions in Liquid Haskell. For example, let us prove that reverse is its own inverse, *i.e.,* reverse (reverse xs) == xs. We express this property as the type of a function involutionP, whose body constitutes the proof:

```
{-@ involutionP :: xs:[a] →
        {reverse (reverse xs) == xs} @-}

involutionP :: [a] → Proof
involutionP []
```

```
   =   reverse (reverse [])
       -- applying inner reverse
   ==. reverse []
       -- applying reverse
   ==. []
   *** QED
involutionP (x:xs)
   =   reverse (reverse (x:xs))
       -- applying inner reverse
   ==. reverse (reverse xs ++ [x])
       ? distributivityP (reverse xs) [x]
   ==. reverse [x] ++ reverse (reverse xs)
       ? involutionP xs
   ==. reverse [x] ++ xs
       ? singletonP x
   ==. [x] ++ xs
       -- applying ++
   ==. x:([] ++ xs)
       -- applying ++
   ==. (x:xs)
   *** QED
```

Because `involutionP` is a recursive function, this constitutes a proof by induction. The two equations for `involutionP` correspond to the two cases of the induction principle:

- In the base case, because the body of the function contains the terms `reverse (reverse [])` and `reverse []`, the corresponding equations are passed to the SMT solver, which then proves that `reverse (reverse []) = []`.

- In the inductive case, we need to show that `reverse (reverse (x:xs)) = (x:xs)`, which proceeds in several steps. The validity of each step is checked by Liquid Haskell when verifying that the refinement type of (==.) is satisfied. Some of the steps follow directly from definitions, and we just add a comment for clarity. Other steps require external lemmas or the inductive hypothesis, which we invoke via the explanation operator (?).

We use the lemma `distributivityP`, which states that list reversal distributes (contravariantly) over list append:

```
{-@ distributivityP :: xs:[a] → ys:[a] →
    {reverse (xs ++ ys)
       == reverse ys ++ reverse xs} @-}
```

Again, we define `distributivityP` as a recursive function, as the property can be proven by induction:

```
distributivityP [] ys
   =   reverse ([] ++ ys)
   ==. reverse ys
       ? rightIdP (reverse ys)
   ==. reverse ys ++ []
   ==. reverse ys ++ reverse []
   *** QED

distributivityP (x:xs) ys
   =   reverse ((x:xs) ++ ys)
   ==. reverse (x:(xs ++ ys))
   ==. reverse (xs ++ ys) ++ [x]
       ? distributivityP xs ys
   ==. (reverse ys ++ reverse xs) ++ [x]
       ? assocP (reverse ys) (reverse xs) [x]
   ==. reverse ys ++ (reverse xs ++ [x])
   ==. reverse ys ++ reverse (x:xs)
   *** QED
```

This proof itself requires additional lemmas about append, namely right identity (`rightIdP`) and associativity (`assocP`), which we tackle with further automation below.

### 2.4 Proof Automation

In the proofs presented so far, we explicitly wrote every step of a function's evaluation. For example, in the base case of `involutionP` we twice applied the function `reverse` to the empty list. Writing proofs explicitly in this way is often helpful (for instance, it makes clear that to prove that `reverse` is an involution we need to prove that it distributes over append) but it can quickly become tedious.

To simplify proofs, Liquid Haskell employs the complete and terminating proof technique of *Proof By (Logical) Evaluation* (PLE) [Vazou et al. 2018]. Conceptually, PLE executes functions for as many steps as needed and automatically provides all the resulting equations to the SMT solver.

Without using this technique, we could prove that the empty list is append's right identity as follows:

```
{-@ rightIdP :: xs:[a] → { xs ++ [] == xs } @-}
rightIdP :: [a] → Proof
rightIdP []
   =   [] ++ []
   ==. []
   *** QED
rightIdP (x:xs)
   =   (x:xs) ++ []
   ==. x : (xs ++ []) ? rightIdP xs
   ==. x : xs
   *** QED
```

However, we can activate PLE in the definition of `rightIdP` using the `ple rightIdP` annotation. This automates all the rewriting steps, and the proof can be simplified to:

```
{-@ rightIdP :: xs:[a] → { xs ++ [] == xs } @-}

{-@ ple rightIdP @-}
rightIdP :: [a] → Proof
rightIdP []     = ()
rightIdP (_:xs) = rightIdP xs
```

That is, the base case is fully automated by PLE, while in the inductive case we must make a recursive call to get the induction hypothesis, but the rest is taken care of by PLE.

Functional Pearl: Theorem Proving for All    5Using this technique we can also prove the remaining lemma, namely the associativity of append:

```
{-@ assocP :: xs:[a] → ys:[a] → zs:[a] →
    {xs ++ (ys ++ zs) == (xs ++ ys) ++ zs}  @-}

{-@ ple assocP @-}
assocP :: [a] → [a] → [a] → Proof
assocP []     _ _ = ()
assocP (_:xs) ys zs = assocP xs ys zs
```

Again, we only have to give the structure of the induction and the arguments to the recursive call, and the PLE machinery adds all the necessary equations to complete the proof.

PLE is a powerful tool that makes proofs shorter and easier to write. However, proofs using this technique are usually more difficult to read, as they hide the details of function expansion. For this reason, while we could apply PLE to simplify many of the proofs in this paper, we prefer to spell out each step. Doing so keeps our proofs easier to understand and in close correspondence with the pen-and-paper proofs we reference in Hutton [2016]'s book.

## 3 Totality and Termination

At this point some readers might be concerned that using a recursive function to model a proof by induction is not sound if the recursive function is partial or non-terminating. However, Liquid Haskell also provides a powerful totality and termination checker and rejects any definition that it cannot prove to be total and terminating.

### 3.1 Totality Checking

Liquid Haskell uses GHC's pattern completion mechanism to ensure that all functions are total. For example, if the `involutionP` was only defined for the empty list case,

```
involutionP :: [a] → Proof
involutionP [] = ()
```

then an error message would be displayed:

```
Your function isn't total:
some patterns aren't defined.
```

To achieve this result, GHC first completes the `involutionP` definition by adding a call to the `patError` function:

```
involutionP [] = ()
involutionP _  = patError "function involutionP"
```

Liquid Haskell then enables totality checking by refining the `patError` function with a false precondition:

```
{-@ patError :: { i:String | False } → a @-}
```

Because there is no argument that satisfies `False`, when calls to the `patError` function cannot be proved to be dead code, Liquid Haskell raises a totality error.

### 3.2 Termination Checking

Liquid Haskell checks that all functions are terminating, using either structural or semantic termination checking.

***Structural Termination***   Structural termination checking is fully automatic and detects the common recursion pattern where the argument to the recursive call is a direct or indirect subterm of the original function argument, as with `length`. If the function has multiple arguments, then at least one argument must get smaller, and all arguments before that must be unchanged (lexicographic order).

In fact, all recursive functions in this paper are accepted by the structural termination checker. This shows that language learners can do a lot before they have to deal with termination proof issues. Eventually, though, they will reach the limits of structural recursion, which is when they can turn to the other technique of semantic termination.

***Semantic Termination***   When the structural termination check fails, Liquid Haskell tries to prove termination using a semantic argument, which requires an explicit termination argument: an expression that calculates a natural number from the function's argument and which decreases in each recursive call. We can use this termination check for the proof `involutionP`, using the syntax `/ [length xs]`:

```
{-@ involutionP :: xs:[a] →
      {reverse (reverse xs) == xs}
      / [length xs] @-}
```

A termination argument has the form `/ [e1, ..., en]`, where the expressions `ei` depend on the function arguments and produce natural numbers. They should lexicographically decrease at each recursive call. These proof obligations are checked by the SMT solver, together with all the refinement types of the function. If the user does not specify a termination metric, but the structural termination check fails, Liquid Haskell tries to guess a termination metric where the first non-function argument is decreasing.

Semantic termination has two main benefits over structural termination: Firstly, not every function is structurally recursive, and making it such by adding additional parameters can be cumbersome and cluttering. And secondly, since termination is checked by the SMT solver, it can make use of refinement properties of the inputs. However, semantic termination also has two main drawbacks. Firstly, when the termination argument is trivial, then the calls to the solver can be expensive. And secondly, termination checking often requires explicitly providing the termination metric, such as the `length` of an input list.

### 3.3 Uncaught termination

Because Haskell is pure, the only effects it allows are divergence and incomplete patterns. If we rule out both these effects, using termination and totality checking, the user can rest assured that their functions are total, and thus correctly encode mathematical proofs.

Unfortunately, creative use of certain features of Haskell, in particular types with non-negative recursion and higher-rank types, can be used to write non-terminating functions that pass Liquid Haskell's current checks. Until this is fixed[3], users need to be careful when using such features.

## 4 Function Optimization

Equational reasoning is not only useful to verify existing code, it can also be used to derive new, more performant function definitions from specifications.

### 4.1 Example: Reversing a List

The `reverse` function that we defined in § 2 was simple and easy to reason about, but it is also rather inefficient. In particular, for each element in the input list, `reverse` appends it to the end of the reversed tail of the list:

```
reverse (x:xs) = reverse xs ++ [x]
```

Because the runtime of `++` is linear in the length of its first argument, the runtime of `reverse` is quadratic. For example, reversing a list of ten thousand elements would take around fifty million reduction steps, which is excessive.

To improve the performance, we would like to define a function that does the reversing and appending *at the same time*; that is, to define a new function

```
reverseApp :: [a] → [a] → [a]
```

that reverses its first argument and appends its second. We can express this as a Liquid Haskell specification:

```
{-@ reverseApp :: xs:[a] → ys:[a] →
      {zs:[a] | zs == reverse xs ++ ys} @-}
```

We now seek to derive an implementation for `reverseApp` that satisfies this specification and is efficient.

***Step 0: Specification***  We begin by writing a definition for `reverseApp` that trivially satisfies the specification and is hence accepted by Liquid Haskell, but is not yet efficient:

```
{-@ reverseApp :: xs:[a] → ys:[a] →
      {zs:[a] | zs == reverse xs ++ ys} @-}
reverseApp :: [a] → [a] → [a]
reverseApp xs ys = reverse xs ++ ys
```

We then seek to improve the definition for `reverseApp` in step-by-step manner, using Liquid Haskell's equational reasoning facilities to make sure that we don't make any mistakes, *i.e.,* that we do not violate the specification.

---

[3]https://github.com/ucsd-progsys/liquidhaskell/issues/159

***Step 1: Case Splitting***  Most likely, the function has to analyse its argument, so let us pattern match on the first argument `xs` and update the right-hand side accordingly:

```
{-@ reverseApp :: xs:[a] → ys:[a] →
      {zs:[a] | zs == reverse xs ++ ys} @-}
reverseApp :: [a] → [a] → [a]
reverseApp []     ys = reverse []     ++ ys
reverseApp (x:xs) ys = reverse (x:xs) ++ ys
```

Liquid Haskell ensures that our pattern match is total, and that we updated the right-hand side correctly.

***Step 2: Equational Rewriting***  Now we seek to rewrite the right-hand sides of `reverseApp` to more efficient forms, while ensuring that our function remains correct. To do so, we can use the `(==.)` operator to show that each change we make gives us the same function. Whenever we add a line, Liquid Haskell will check that this step is valid. We begin by simply expanding definitions:

```
{-@ reverseApp :: xs:[a] → ys:[a] →
      {zs:[a] | zs == reverse xs ++ ys} @-}
reverseApp :: [a] → [a] → [a]
reverseApp [] ys
  =   reverse [] ++ ys
  ==. [] ++ ys
  ==. ys
reverseApp (x:xs) ys
  =   reverse (x:xs) ++ ys
  ==. (reverse xs ++ [x]) ++ ys
```

At this point, we have expanded as much as we can, but `reverseApp` still uses the original, inefficient `reverse` functions, so we are not done. However, we proved at the end of § 2 that append is associative, so we can use this fact to transform `(reverse xs ++ [x]) ++ ys` into `reverse xs ++ ([x] ++ ys)`, and then continue expanding:

```
  ...
  ==. (reverse xs ++ [x]) ++ ys
      ? assocP (reverse xs) [x] ys
  ==. reverse xs ++ ([x] ++ ys)
  ==. reverse xs ++ (x:([] ++ ys))
  ==. reverse xs ++ (x:ys)
```

We're still using `reverse`, so we're not quite done. To finish the definition, we just need to observe that the last line has the form `reverse as ++ bs` for some lists as and bs. This is precisely the form of the specification for `reverseApp as bs`, so we can rewrite the last line in terms of `reverseApp`:

```
  ...
  ==. reverse xs ++ (x:ys)
  ==. reverseApp xs (x:ys)
```

In summary, our definition for `reverseApp` no longer mentions the `reverse` function or the append operator. Instead, it contains a recursive call to `reverseApp`, which means we have derived the following, self-contained definition:

```
reverseApp :: [a] → [a] → [a]
reverseApp []     ys = ys
reverseApp (x:xs) ys = reverseApp xs (x:ys)
```

The runtime performance of this definition is linear in the length of its first argument, a significant improvement.

***Step 3: Elimination of Equational Steps*** We can obtain the small, self-contained definition for reverseApp by deleting all lines but the last from each case of the derivation. But we do not have to! Recall that the (==.) operator is defined to simply return its second argument. So semantically, both definitions of reverseApp are equivalent.

One might worry that all the calls to reverse, reverseApp, (++) and assocP in the derivation will spoil the performance of the function, but because Haskell is a lazy language, in practice none of these calls are actually executed. And in fact (with optimizations turned on), the compiler completely removes them from the code and — as we confirmed using inspection testing [Breitner 2018] — both definitions of reverseApp optimize to identical intermediate code.

In conclusion, we can happily leave the full derivation in the source file and obtain precisely the same performance as if we had used the self-contained definition for reverseApp given at the end of the previous step.

***Step 4: Optimizing reverse*** The goal of this exercise was not to have an efficient reverse-and-append function, but to have an efficient reverse function. However, we can define this using reverseApp, again starting from its specification and deriving the code that we want to run. Here we need to turn reverse xs into reverse xs ++ ys for some list ys. This requires us to use the theorem rightIdP:

```
{-@ reverse' :: xs:[a] →
      {v:[a] | v == reverse xs } @-}
reverse' :: [a] → [a]
reverse' xs
  =    reverse xs ? rightIdP (reverse xs)
  ==.  reverse xs ++ []
  ==.  reverseApp xs []
```

The above derivation follows the same steps as the pen-and-paper version in [Hutton 2016], with one key difference: the correctness of each step, and the derived program, is now automatically checked by Liquid Haskell.

### 4.2 Example: Flattening a Tree

We can use the same technique to derive an optimized function for flattening trees. Our trees are binary trees with integers in the leaves, as in [Hutton 2016]:

```
data Tree = Leaf Int | Node Tree Tree
```

We wish to define an efficient function that flattens such a tree to a list. As with reverse, we begin with a simple but inefficient version that uses the append operator:

```
{-@ reflect flatten @-}
flatten :: Tree → [Int]
flatten (Leaf n)   = [n]
flatten (Node l r) = flatten l ++ flatten r
```

Because we want to refer to this function in our specifications and reasoning, we instruct Liquid Haskell to lift it to the refinement type level using **reflect** keyword. Liquid Haskell's structural termination checker (§ 3.2) accepts this definition and all following functions on trees, and there is no need to define a measure on trees.

We can use flatten as the basis of a specification for a more efficient version. As before, the trick is to combine flatten with list appending and define a function

```
flattenApp :: Tree → [Int] → [Int]
```

with the specification flattenApp t ns == flatten t ++ ns, which we can state as a Liquid Haskell type signature:

```
{-@ flattenApp :: t:Tree → ns:[Int] →
      {v:[Int] | v == flatten t ++ ns } @-}
```

As in the previous example, we begin by using the specification as a correct but inefficient implementation

```
flattenApp t ns = flatten t ++ ns
```

and use equational reasoning in Liquid Haskell to work our way towards an implementation that avoids the use of the inefficient flatten and append functions:

```
flattenApp :: Tree → [Int] → [Int]
flattenApp (Leaf n) ns
  =    flatten (Leaf n) ++ ns
  ==.  [n] ++ ns
  ==.  n:([] ++ ns)
  ==.  n:ns
flattenApp (Node l r) ns
  =    flatten (Node l r) ++ ns
  ==.  (flatten l ++ flatten r) ++ ns
       ? assocP (flatten l) (flatten r) ns
  ==.  flatten l ++ (flatten r ++ ns)
  ==.  flatten l ++ (flattenApp r ns)
  ==.  flattenApp l (flattenApp r ns)
```

Again, this derivation serves both as an implementation and a verification, and is operationally equivalent to:

```
flattenApp :: Tree → [Int] → [Int]
flattenApp (Leaf n)   ns = n:ns
flattenApp (Node l r) ns =
  flattenApp l (flattenApp r ns)
```

Finally, we can then derive the optimized flatten function by means of the following simple reasoning:

```
{-@ flatten' :: t:Tree →
      {v:[Int] | v == flatten t} @-}
flatten' :: Tree → [Int]
flatten' l
  =    flatten l ? rightIdP (flatten l)
```

8Niki Vazou, Joachim Breitner, Will Kunkel, David Van Horn, Graham Hutton

```
      ==. flatten l ++ []
      ==. flattenApp l []
```

In conclusion, the derivation once again follows the same steps as the original pen-and-paper version, but is now mechanically checked for correctness.

## 5   Case Study: Correct Compilers

So far, all the proofs that we have seen have been very simple. To show that Liquid Haskell scales to more involved arguments, we show how it can be used to calculate a correct and efficient compiler for arithmetic expressions with addition, as in [Bahr and Hutton 2015; Hutton 2016].

We begin by defining an expression as an integer value or the addition of two expressions, and a function that returns the integer value of such an expression:

```
data Expr = Val Int | Add Expr Expr

{-@ reflect eval @-}
eval :: Expr → Int
eval (Val n)   = n
eval (Add x y) = eval x + eval y
```

***A simple stack machine***   The target for our compiler will be a simple stack-based machine. In this setting, a stack is a list of integers, and code for the machine is a list of push and add operations that manipulate the stack:

```
type Stack = [Int]
type Code  = [Op]
data Op    = PUSH Int | ADD
```

The meaning of such code is given by a function that executes a piece of code on an initial stack to give a final stack:

```
{-@ reflect exec @-}
exec :: Code → Stack → Stack
exec []         s       = s
exec (PUSH n:c) s       = exec c (n:s)
exec (ADD:c)    (m:n:s) = exec c (n+m:s)
```

That is, PUSH places a new integer on the top of the stack, while ADD replaces the top two integers by their sum.

***A note on totality***   The function exec is not total — in particular, the result of executing an ADD operation on a stack with fewer than two elements is undefined. Like most proof systems, Liquid Haskell requires all functions to be total in order to preserve soundness. There are a number of ways we can get around this problem, such as:

- Using Haskell's Maybe type to express the possibility of failure directly in the type of the exec function.
- Adding a refinement to exec to specify that it can only be used with "valid" code and stack arguments.
- Arbitrarily defining how ADD operates on a small stack, for example by making it a no-operation.

- Using dependent types to specify the stack demands of each operation in our language [Mckinna and Wright 2006]. For example, we could specify that ADD transforms a stack of length $n + 2$ to a stack of length $n + 1$.

For simplicity, we adopt the first approach here, and rewrite exec as a total function that returns Nothing in the case of failure, and Just s in the case of success:

```
exec :: Code → Stack → Maybe Stack
exec []         s       = Just s
exec (PUSH n:c) s       = exec c (n:s)
exec (ADD:c)    (m:n:s) = exec c (n+m:s)
exec _          _       = Nothing
```

***Compilation***   We now want to define a compiler from expressions to code. The property that we want to ensure is that executing the resulting code will leave the value of the expression on top of the stack. Using this property, it is clear that an integer value should be compiled to code that simply pushes the value onto the stack, while addition can be compiled by first compiling the two argument expressions, and then adding the resulting two values:

```
{-@ reflect comp @-}
comp :: Expr → Code
comp (Val n)   = [PUSH n]
comp (Add x y) = comp x ++ comp y ++ [ADD]
```

Note that when an add operation is performed, the value of the expression y will be on top of the stack and the value of expression x will be below it, hence the swapping of these two values in the definition of the exec function.

***Correctness***   The correctness of the compiler for expressions is expressed by the following equation:

```
exec (comp e) [] == Just [eval e]
```

That is, compiling an expression and executing the resulting code on an empty stack always succeeds, and leaves the value of the expression as the single item on the stack. In order to prove this result, however, we will find that it is necessary to generalize to an arbitrary initial stack:

```
exec (comp e) s == Just (eval e : s)
```

We prove correctness of the compiler in Liquid Haskell by defining a function generalizedCorrectnessP with a refinement type specification that encodes the above equation. We define the body of this function by recursion on the type Expr, which is similar to induction for the type Tree in § 4.2. We begin as before by expanding definitions:

```
{-@ generalizedCorrectnessP
    :: e:Expr → s:Stack
    → {exec (comp e) s == Just (eval e:s)} @-}
generalizedCorrectnessP
    :: Expr → Stack → Proof
generalizedCorrectnessP (Val n) s
  =   exec (comp (Val n)) s
```

```
    ==. exec [PUSH n] s
    ==. exec [] (n:s)
    ==. Just (n:s)
    ==. Just (eval (Val n):s)
    *** QED

generalizedCorrectnessP (Add x y) s
   =   exec (comp (Add x y)) s
   ==. exec (comp x ++ comp y ++ [ADD]) s
   ...
```

That is, we complete the proof for Val by simply expanding definitions, while for Add we quickly reach a point where we need to think further. Intuitively, we require a lemma which states that executing code of the form c ++ d would give the same result as executing c and then executing d:

```
exec (c ++ d) s == exec d (exec c s)
```

Unfortunately, this doesn't typecheck, because exec takes a Stack but returns a Maybe Stack. What we need is some way to run exec d only if exec c succeeds. Fortunately, this already exists in Haskell — it's just monadic bind for the Maybe type, which we reflect in Liquid Haskell as follows:

```
{-@ reflect >>= @-}
(>>=) :: Maybe a → (a → Maybe b) → Maybe b
(Just x) >>= f = f x
Nothing  >>= _ = Nothing
```

We can now express our desired lemma using bind

```
exec (c ++ d) s == exec c s >>= exec d
```

and its proof proceeds by straightforward structural induction on the first code argument, with separate cases for success and failure of an addition operator:

```
{-@ sequenceP :: c:Code → d:Code → s:Stack →
    {exec (c ++ d) s == exec c s >>= exec d} @-}
sequenceP :: Code → Code → Stack → Proof
sequenceP [] d s
   =   exec ([] ++ d) s
   ==. exec d s
   ==. Just s >>= exec d
   ==. exec [] s >>= exec d
   *** QED

sequenceP (PUSH n:c) d s
   =   exec ((PUSH n:c) ++ d) s
   ==. exec (PUSH n:(c ++ d)) s
   ==. exec (c ++ d) (n:s)
       ? sequenceP c d (n:s)
   ==. exec c (n:s) >>= exec d
   ==. exec (PUSH n:c) s >>= exec d
   *** QED

sequenceP (ADD:c) d (m:n:s)
   =   exec ((ADD:c) ++ d) (m:n:s)
   ==. exec (ADD:(c ++ d)) (m:n:s)
   ==. exec (c ++ d) (n + m:s)
       ? sequenceP c d (n + m:s)
   ==. exec c (n + m:s) >>= exec d
   ==. exec (ADD:c) (m:n:s) >>= exec d
   *** QED

sequenceP (ADD:c) d s
   =   exec ((ADD:c) ++ d) s
   ==. exec (ADD:(c ++ d)) s
   ==. Nothing
   ==. Nothing >>= exec d
   ==. exec (ADD:c) s >>= exec d
   *** QED
```

With this lemma in hand, we can complete the Add case of our generalizedCorrectnessP proof:

```
...
generalizedCorrectnessP (Add x y) s
   =   exec (comp (Add x y)) s
   ==. exec (comp x ++ comp y ++ [ADD]) s
       ? sequenceP (comp x) (comp y ++ [ADD]) s
   ==. exec (comp x) s >>= exec (comp y ++ [ADD])
       ? generalizedCorrectnessP x s
   ==. Just (eval x:s) >>= exec (comp y ++ [ADD])
   ==. exec (comp y ++ [ADD]) (eval x:s)
       ? sequenceP (comp y) [ADD] (eval x:s)
   ==. exec (comp y) (eval x:s) >>= exec [ADD]
       ? generalizedCorrectnessP y (eval x:s)
   ==. Just (eval y:eval x:s) >>= exec [ADD]
   ==. exec [ADD] (eval y:eval x:s)
   ==. exec [] (eval x + eval y:s)
   ==. Just (eval x + eval y:s)
   ==. Just (eval (Add x y):s)
   *** QED
```

Now that we have proven a generalized version of our correctness theorem, we can recover the original theorem by replacing the arbitrary state s by the empty state []:

```
{-@ correctnessP :: e:Expr →
      {exec (comp e) [] == Just [eval e]} @-}
correctnessP :: Expr → Proof
correctnessP e = generalizedCorrectnessP e []
```

***A faster compiler*** Notice that like reverse and flatten, our compiler uses the append operator (++) in the recursive case. This means that our compiler can be optimized. We can use the same strategy as we used for reverse and flatten to derive an optimized version of comp.

We begin by defining a function compApp with the property compApp e c == comp e ++ c. As previously, we proceed from this property by expanding definitions and applying lemmata to obtain an optimized version:

```
{-@ reflect compApp @-}
```

```
{-@ compApp :: e:Expr → c:Code →
    {d:Code | d == comp e ++ c} @-}
compApp (Val n) c
  =   comp (Val n) ++ c
  ==. [PUSH n] ++ c
  ==. PUSH n:([] ++ c)
  ==. PUSH n:c

compApp (Add x y) c
  =   comp (Add x y) ++ c
  ==. (comp x ++ comp y ++ [ADD]) ++ c
      ? appAssocP (comp x) (comp y ++ [ADD]) c
  ==. comp x ++ (comp y ++ [ADD]) ++ c
      ? appAssocP (comp y) [ADD] c
  ==. comp x ++ comp y ++ [ADD] ++ c
  ==. comp x ++ comp y ++ ADD:([] ++ c)
  ==. comp x ++ comp y ++ ADD:c
  ==. comp x ++ compApp y (ADD:c)
  ==. compApp x (compApp y (ADD:c))
```

The Haskell compiler automatically optimizes away all the equational reasoning steps to derive the following definition for compApp, which no longer makes use of append:

```
compApp :: Expr → Code → Code
compApp (Val n)   c = PUSH n:c
compApp (Add x y) c =
    compApp x (compApp y (ADD:c))
```

From this definition, we can construct the optimized compiler by supplying the empty list as the second argument:

```
{-@ reflect comp' @-}
comp' :: Expr → Code
comp' e = compApp e []
```

In turn, we can then prove that new compiler comp' is equivalent to the original version comp, and is hence correct:

```
{-@ equivP :: e:Expr → {comp' e == comp e} @-}
equivP e
  =   comp' e
  ==. compApp e []
  ==. comp e ++ [] ? appRightIdP (comp e)
  ==. comp e
  *** QED

{-@ equivCorrectnessP :: e:Expr →
    {exec (comp' e) [] == Just [eval e]} @-}
equivCorrectnessP e =
      exec (comp' e) []    ? equivP e
  ==. exec (comp e) [] ? correctnessP e
  ==. Just [eval e]
  *** QED
```

However, we can also prove the correctness of comp' without using comp at all — and it turns out that this proof is much simpler. Again, we generalize our statement of correctness, this time to any initial stack and any additional code:

```
exec (compApp e c) s == exec c (cons (eval e) s)
```

We can then prove this new correctness theorem by induction on the structure of the expression argument:

```
{-@ generalizedCorrectnessP'
      :: e:Expr → s:Stack → c:Code →
         { exec (compApp e c) s ==
           exec c (cons (eval e) s)} @-}
generalizedCorrectnessP'
      :: Expr → Stack → Code → Proof
generalizedCorrectnessP' (Val n) s c
  =   exec (compApp (Val n) c) s
  ==. exec (PUSH n:c) s
  ==. exec c (n:s)
  ==. exec c (eval (Val n):s)
  *** QED

generalizedCorrectnessP' (Add x y) s c
  =   exec (compApp (Add x y) c) s
  ==. exec (compApp x (compApp y (ADD:c))) s
      ? generalizedCorrectnessP' x s
        (compApp y (ADD:c))
  ==. exec (compApp y (ADD:c)) (eval x:s)
      ? generalizedCorrectnessP' y (eval x:s)
        (ADD:c)
  ==. exec (ADD:c) (eval y:eval x:s)
  ==. exec c (eval x + eval y:s)
  ==. exec c (eval (Add x y):s)
  *** QED
```

Finally, we recover our original correctness theorem by specializing both the stack s and code c to empty lists:

```
{-@ correctnessP' :: e:Expr →
    {exec (comp' e) [] == Just [eval e]} @-}
correctnessP' :: Expr → Proof
correctnessP' e
  =   exec (comp' e) []
  ==. exec (compApp e []) []
      ? generalizedCorrectnessP' e [] []
  ==. exec [] [eval e]
  ==. Just [eval e]
  *** QED
```

In summary, there are two key benefits to our new compiler. First of all, it no longer uses append, and is hence more efficient. And secondly, its correctness proof no longer requires the sequenceP lemma, and is hence simpler and more concise. Counterintuitively, code optimized using Liquid Haskell can be easier to prove correct, not harder!

## 6 Related Work

The equational reasoning in this article takes the form of inductive proofs about terminating Haskell functions, so it is possible to reproduce the proofs in most general-purpose theorem provers. Below we compare a number of such theorem provers with the use of Liquid Haskell.

***Coq*** The Coq system [Bertot and Castéran 2004] is the prototypical example of a theorem prover based on dependent types. For example, we can use Coq to prove that the list append operator is associative [Pierce et al. 2018]:

```
Theorem app_assoc : forall l1 l2 l3 : natlist,
  (l1 ++ l2) ++ l3 = l1 ++ (l2 ++ l3).
Proof.
  intros l1 l2 l3.
  induction l1 as [| n l1' IHl1'].
  - reflexivity.
  - simpl. rewrite -> IHl1'. reflexivity.
Qed.
```

This proof resembles the PLE-enabled version of the Liquid Haskell proof presented in § 2. However, while the proof in Liquid Haskell can be easily expanded to show all steps in an equational-reasoning style, there is no straightforward way to do the same in the Coq system.

Moreover, to reason about Haskell code in Coq, we must first translate our Haskell code into Coq's programming language, Gallina. While the recently developed tool hs-to-coq [Spector-Zabusky et al. 2018] can automate this translation, Haskell programmers still have to learn Coq to actually prove results about the translated code. In contrast, equational reasoning in Liquid Haskell allows the users to reason about their Haskell code while expressing their proofs using Haskell's familiar syntax and semantics.

Coq is not without its benefits [Vazou et al. 2017], however. It provides an extensive library of theorems, interactive proof assistance, and a small trusted code base, all of which are currently lacking in Liquid Haskell. Of course, these benefits are not limited to Coq, and we could also port our code, theorems and proofs to other dependently-typed languages, such as Idris [Brady 2013] and F* [Swamy et al. 2016].

***Isabelle*** We could also use theorem provers based on higher-order logic, such as Isabelle [Nipkow et al. 2002]. Isabelle's powerful automation based on term rewriting requires us to merely indicate the right induction principle:

```
lemma app_assoc:
  "(xs ++ ys) ++ zs = xs ++ (ys ++ zs)"
  by (induction xs) auto
```

This proofs resembles the concise PLE version of the Liquid Haskell proof. Isabelle's language for declarative proofs, Isar [Nipkow 2002], supports equational reasoning, and also permits spelling out the proof in full detail:

```
lemma app_assoc:
  "(xs ++ ys) ++ zs = xs ++ (ys ++ zs)"
proof(induction xs)
  case Nil
    have "([] ++ ys) ++ zs = ys ++ zs" by simp
    also have "... = [] ++ (ys ++ zs)" by simp
    finally show ?case.
  next
  case (Cons x xs)
    have "(x:xs ++ ys) ++ zs = x:(xs ++ ys) ++ zs"
      by simp
    also have "... = x:xs ++ (ys ++ zs)"
      by (simp add: Cons.IH)
    also have "... = (x:xs) ++ (ys ++ zs)" by simp
    finally show ?case.
qed
```

Each equational step is verified by the Isabelle system using its term rewriting engine simp, and the use of the inductive hypothesis (Cons.IH) is clearly marked.

The tool Haskabelle can translate Haskell function definitions into Isabelle [Haftmann 2010].

***Other theorem provers*** Support for equational reasoning in this style is also built into Lean [de Moura et al. 2015], a general semi-automated theorem prover, and Dafny [Leino 2010], an imperative programming language with built-in support for specification and verification using SMT solvers [Leino and Polikarpova 2013].

### 6.1 Operator-Based Equational Reasoning

The support for equational reasoning in Isabelle, Lean and Dafny is built into their syntax, while in Liquid Haskell, the operators for equational reasoning are provided by a library. This approach is highly inspired by Agda.

Agda [Norell 2007] is a general theorem prover based on dependent type theory. Its type system and syntax is flexible enough to allow the library-defined operator

$$\_\equiv\langle\_\rangle\_\ :\ \forall\ (x\ \{y\ z\}\ :\ A)\ \to\ x\equiv y\ \to\ y\equiv z\ \to\ x\equiv z$$

which expresses an equality together with its proof, and is similar to Liquid Haskell's (==.) operator:

```
a ≡⟨ explanation ⟩ b   -- Agda
a ==. b ? explanation  -- Liquid Haskell
```

One disadvantages of the operator-based equational reasoning in Liquid Haskell over built-in support as provided in, say, Dafny is that there each equation is checked independently, whereas in Liquid Haskell all equalities in one function are checked at once, which can be slower.

While the above tools support proofs using equational reasoning, Liquid Haskell is unique in extending an existing, general-purpose programming language to support theorem proving. This makes Liquid Haskell a more natural choice for verifying Haskell code, both because it is familiar to Haskell programmers, and because it does not require porting code to a separate verification language.

### 6.2 Verification in Haskell

Liquid Haskell is not the first attempt to bring theorem proving to Haskell. Zeno [Sonnex et al. 2012] generates proofs by term rewriting and Halo [Vytiniotis et al. 2013] uses an axiomatic encoding to verify contracts. Both these tools are automatic, but unpredictable and not programmer-extensible, which has limited them to verifying much simpler properties than the ones checked here. Another tool, HERMIT [Farmer et al. 2015], proves equalities by rewriting the GHC core language, guided by user specified scripts. Compared to these tools, in Liquid Haskell the proofs are Haskell programs while SMT solvers are used to automate reasoning.

Haskell itself now supports dependent types [Eisenberg 2016], where inductive proofs can be encoded as type class constraints. However, proofs in Dependent Haskell do not have the straightforward equational-reasoning style that Liquid Haskell allows and are not SMT-aided.

## 7 Conclusion

We demonstrated how Hutton [2016]'s equational reasoning proofs can be encoded as Haskell programs and be checked using Liquid Haskell. The proofs include equational reasoning for functional correctness, program optimization and program calculation. The encoding from pen-and-paper proofs into machine checked proofs is direct, thus we claim that Liquid Haskell is a theorem prover that can be naturally used by any Haskell programmer or learner.